\documentclass[twocolumn]{aastex61}
\usepackage{amsmath}
\usepackage{amsfonts}
\usepackage{amssymb}
\usepackage{gensymb}
\usepackage{float}
\usepackage{graphicx}
\usepackage{calc}
\usepackage[toc,page]{appendix}
\usepackage[section]{placeins}
\usepackage{morefloats}
\usepackage{mathrsfs}
\usepackage{amsfonts}
\usepackage{color}
\usepackage{natbib}
\usepackage{longtable}
\newlength{\figwidth}
\newcommand{\harm}{{\tt harm}}
\newcommand{\grmonty}{{\tt grmonty}}
\newcommand{\bhlight}{{\tt bhlight}}
\newcommand{\ebhlight}{{\tt ebhlight}}

\newcommand{\mdot}{$\dot{m}$}

\shorttitle{AASTeX 6.1 Template}
\shortauthors{Ryan et al.}

\begin{document}
%\showthe\columnwidth

\title{Two-Temperature GRRMHD Simulations of M87}

\author{Benjamin R. Ryan}
\affiliation{CCS-2, Los Alamos National Laboratory, P.O. Box 1663, Los Alamos, NM 87545, USA}
\affiliation{Center for Theoretical Astrophysics, Los Alamos National Laboratory, Los Alamos, NM, 87545, USA}
\author{Sean M. Ressler}
\affiliation{Departments of Astronomy \& Physics, Theoretical Astrophysics Center, University of California, Berkeley, CA 94720, USA}
\author{Joshua C. Dolence}
\affiliation{CCS-2, Los Alamos National Laboratory, P.O. Box 1663, Los Alamos, NM 87545, USA}
\affiliation{Center for Theoretical Astrophysics, Los Alamos National Laboratory, Los Alamos, NM, 87545, USA}
\author{Charles Gammie}
\affil{Department of Astronomy, University of Illinois, 1002 West Green Street, Urbana, IL, 61801, USA}
\affil{Department of Physics, University of Illinois, 1110 West Green Street, Urbana, IL, 61801, USA}
\author{Eliot Quataert}
\affiliation{Departments of Astronomy \& Physics, Theoretical Astrophysics Center, University of California, Berkeley, CA 94720, USA}

\begin{abstract}

We present axisymmetric two-temperature general relativistic radiation magnetohydrodynamic (GRRMHD) simulations of the inner region of the accretion flow onto the supermassive black hole M87. We address uncertainties from previous modeling efforts through inclusion of models for (1) self-consistent dissipative and Coulomb electron heating (2) radiation transport (3) frequency-dependent synchrotron emission, self-absorption, and Compton scattering. We adopt a distance $D=16.7$ Mpc, an observer angle $\theta = 20 \degree$, and consider black hole masses $M/M_{\odot} = (3.3\times10^{9}, 6.2\times10^{9})$ and spins $a_{\star} = (0.5, 0.9375)$ in a four-simulation suite. For each $(M, a_{\star})$, we identify the accretion rate that recovers the 230 GHz flux from very long  baseline interferometry measurements. 
We report on disk thermodynamics at these accretion rates ($\dot{M}/\dot{M}_{\mathrm{Edd}} \sim 10^{-5}$). The disk remains geometrically thick; cooling does not lead to a thin disk component. While electron heating is dominated by Coulomb rather than dissipation for $r \gtrsim 10 GM/c^2$, the accretion disk remains two-temperature. Radiative cooling of electrons is not negligible, especially for $r \lesssim 10 GM/c^2$. The Compton $y$ parameter is of order unity.
We then compare derived and observed or inferred spectra, millimeter images, and jet powers. Simulations with $M/M_{\odot} = 3.3\times10^{9}$ are in conflict with observations. These simulations produce millimeter images that are too small, while the low-spin simulation also overproduces X-rays. For $M/M_{\odot} = 6.2\times10^{9}$, both simulations agree with constraints on radio/IR/X-ray fluxes and millimeter image sizes. Simulation jet power is a factor $10^2-10^3$ below inferred values, a possible consequence of the modest net magnetic flux in our models.

\end{abstract}

\section{Introduction} 
The supermassive black hole (SMBH) at the center of the massive elliptical galaxy M87, hereafter simply M87, has been a classic observational target from the millimeter to the $\gamma$-ray for decades. M87 is a valuable laboratory for studying radiatively inefficient accretion flows (RIAF; \citealt{Ichimaru1977, NarayanYi1994, YuanNarayan2014}), jet launching, and the phenomenology of low-luminosity active galactic nuclei (LLAGN), which dominate the population of local SMBHs (\citealt{GreeneHo2007}).

% M87 mass and spin
Apart from details of the accretion disk, the appearance of a black hole is set by its mass $M$ and dimensionless spin parameter $a_{\star}$. The two leading methods for determining the mass of M87, stellar-dynamical measurements (e.g.\ \citealt{Gebhardt+2011}, who find $M = 6.6\times10^9 M_{\odot}$ for distance $D=17.9$ Mpc) and gas-dynamical measurements (e.g.\ \citealt{Walsh+2013}, who find $M = 3.5\times10^9 M_{\odot}$, also for $D=17.9$ Mpc) currently disagree by a factor $\approx 2$. Note that in this work we prefer $D=16.7$ Mpc (\citealt{Blakeslee+2009}). The spin of M87 is uncertain (although see \citealt{Doeleman+2012} for an argument based on very long baseline interferometry (VLBI) favoring $a_{\star} \gtrsim 0.5$).

% Spectral observations
M87 is detectable at essentially all observed wavelengths: the radio (e.g.\ \citealt{Hada+2011, de Gasperin+2012, Doeleman+2012}), IR (e.g.\ \citealt{Shi+2007, Asmus+2014}), optical/UV (e.g.\ \citealt{Sparks+1996}), X-ray (e.g.\ \citealt{Bohringer+2001, WilsonYang2002, Di Matteo+2003}), and $\gamma$-ray (e.g.\ \citealt{Abdo+2009, Abramowski+2012}). Constructing broadband spectra of LLAGN, however, leads to difficulties: (1) different frequency bands use different observational techniques, leading to inconsistent aperture sizes (2) LLAGN exhibit variability, often on the timescale at which observations at different frequencies may be performed (3) The jet of M87 exhibits several bright knots, especially HST-1 (e.g.\ \citealt{Perlman+2011}). \cite{Prieto+2016} have addressed these issues for M87, creating an optimal set of contemporaneous measurements that led to the identification of two states of accretion: quiescence and outburst. In both cases, the spectrum is nearly flat and featureless across almost 10 decades in frequency, in contrast to typical RIAF models, which contain distinct Compton bumps, at least for a thermal electron distribution function (e.g.\ \citealt{Narayan+1998, Moscibrodzka+2009}).

% EHT
Along with Sagittarius A* (Sgr A*), the Milky Way's SMBH, M87 is one of the two event horizons sufficiently large on the sky for resolved VLBI imaging by the Event Horizon Telescope (EHT, e.g.\ \citealt{Doeleman+2012}). Sgr A* and M87 form a serendipitous pair for studying RIAFs. Despite masses and accretion rates (in Eddington units) differing by several orders of magnitude, and Sgr A* possibly being nearly edge-on (e.g.\ \citealt{Moscibrodzka+2009,Dexter+2010,Shcherbakov+2012}) while M87 is nearly face-on (e.g.\ \citealt{HeinzBegelman1997}), the two sources have approximately the same synchrotron peak frequency. Although Sgr A*'s event horizon is somewhat larger on the sky, particularly if lower measurements for the M87 mass are correct, M87 remains an attractive target for two reasons: (1) intrinsic variability is long compared to the timescale of a global VLBI observation (2) there is only modest interstellar scattering between Earth and M87, in contrast to Sgr A* (\citealt{Bower+2006}). Radio VLBI observations of M87 have already achieved beam sizes of the order of a few Schwarzschild radii (\citealt{Doeleman+2012}), implying a compact population of hot electrons near the black hole, in agreement with previous RIAF models (e.g.\ \citealt{Esin+1997}, \citealt{Moscibrodzka+2009, YuanNarayan2014}).

% MRI and GRMHD
Accretion onto black holes is probably mediated at least in part by angular momentum transport due to the turbulent state resulting from the saturation of the magnetorotational instability (MRI; \citealt{BalbusHawley1991}). 
The magnetic field may also generate long-range correlations in the accretion disk (\citealt{GuanGammie2011}) and produce jets (\citealt{BlandfordZnajek1977}).
These features strongly motivate global general relativistic magnetohydrodynamic (MHD) models of accretion.
As an example of the importance of general relativity for M87, \cite{Dexter+2012} have argued that emission is counterjet dominated through gravitational lensing, a purely relativistic effect.
Significant progress in modeling RIAFs has been made through numerical simulations (e.g.\ \citealt{DeVilliers+2003, McKinneyGammie2004, Narayan+2012, White+2016}), which allow for a self-consistent treatment of the turbulent stress, as well as capturing the effects of large-scale components of the magnetic field (\citealt{Tchekhovskoy+2011, McKinney+2012}).

% Two temp, Coulomb
At very low accretion rates, $\dot{m} \equiv \dot{M}/\dot{M}_{\mathrm{Edd}} \ll 1$ (where the Eddington rate $\dot{M}_{\mathrm{Edd}}$ $\equiv 2.2\times10^{-8}$ $(M/M_{\odot})~M_{\odot}~\mathrm{yr}^{-1}$, i.e.\ we adopt a nominal efficiency $\eta = 0.1$), RIAFs are Coulomb collisionless \citep{MahadevanQuataert1997, Ryan+2017, Sadowski+2017}. Even for such collisionless flows, simple fluid model closures may be sufficient to accurately evolve the total fluid \citep{Foucart+2017}. However, the electron thermodynamics are probably set by Larmor-scale heating and velocity space instabilities (\citealt{Quataert1998, SironiNarayan2015}) which are not captured in ideal MHD. Magnetic reconnection may also play a role in electron heating (\citealt{Rowan+2017}), as well as accelerating nonthermal electrons (e.g.\ \citealt{SironiSpitkovsky2014}) which may have observational consequences for infrared variability and low-frequency radio emission (\citealt{Ozel+2000, Yuan+2003, Chael+2017}).

The generic consequence of electron heating through kinetic turbulent dissipation is probably hot protons and somewhat cooler electrons (\citealt{Quataert1998}). Despite the absence of Coulomb collisions, each population may nonetheless be approximately thermal due to kinetic instabilities that feed off distribution function anisotropies, particularly at higher $\beta \equiv 8 \pi n k_B T / B^2$ (\citealt{Kunz+2014, Riquelme+2015, Kunz+2016}). The electron heating probably depends on the local plasma conditions (\citealt{Howes2010}). While the electron temperature in RIAF simulations is often set to a prescribed fraction of the total internal energy (\citealt{Moscibrodzka+2009, Drappeau+2013, Chan+2015, Moscibrodzka+2016}), \cite{Ressler+2015} have developed a method to combine advection and heating based on implicit dissipation in numerical general relativistic magnetohydrodynamic (GRMHD) schemes to self-consistently evolve the electron temperature, which we extended to include Coulomb coupling in \cite{Ryan+2017} (see also \citealt{Sadowski+2017} for a similar method).

% Sims and observations
Post-processing of nonradiative GRMHD simulations is now a standard technique for interpreting LLAGN observations, particularly for Sgr A* where the accretion rate is so low that radiative feedback on the flow dynamics and energetics is negligible (\citealt{Dibi+2012}). GRMHD models, electron physics, accretion rate, black hole spin, and observer angle are all constrained through spectra, variability, polarization, and imaging (e.g.\ \citealt{Moscibrodzka+2009, Dexter+2012, Dolence+2012, Shcherbakov+2012, Drappeau+2013, ShcherbakovMcKinney2013, Moscibrodzka+2014, Chan+2015, Ball+2016, Medeiros+2017, Ressler+2017}).

% Radiative cooling
A significant challenge to numerical models of M87 has been the apparent importance of radiative processes to the thermodynamics of the accretion flow. Previous efforts applying nonradiative GRMHD simulations to M87 have had difficulty demonstrating self-consistency \citep{Moscibrodzka+2011, Dexter+2012}. Near the black hole, both synchrotron emission and Compton upscattering cool the electrons. More recently, \cite{Moscibrodzka+2016} achieved reasonable radiative efficiencies, but required a proton-to-electron temperature ratio $T_p/T_e=100$ in the midplane, relatively high compared to those preferred for unambiguously nonradiative models (e.g.\ Sgr A*, \citealt{Moscibrodzka+2014}); for $T_p/T_e < 40$ in the midplane, the M87 models overproduced X-ray emission. While optically thin synchrotron emission is easily incorporated, the Compton $y$ parameter is probably $\sim 1$ for M87 (e.g.\ \citealt{Dexter+2012}). Compton scattering globally couples the disk electrons through the radiative transfer equation. Global GRMHD models with self-consistent radiation transport are therefore strongly motivated.

% History of GRRMHD
The importance of radiative cooling to electron temperatures and observable radiation in RIAFs above some $\dot{m}$ has long been recognized (e.g.\ \citealt{Esin+1997, Xie+2010, Niedzwiecki+2012}). However, computational expense and algorithmic complexity have restricted the inclusion of radiative transport into GRMHD calculations. \cite{Ohsuga+2009} and \cite{OhsugaMineshige2011} studied the first global radiation MHD models of accretion disks, using a diffusion model for radiation transport to demonstrate the anticipated transition from RIAFs to radiation-dominated thin disks (\citealt{ShakuraSunyaev1973}) with increasing $\dot{m}$. Subsequent work used local models for radiative cooling (\citealt{FragileMeier2009, Dibi+2012, Wu+2016}), or a fluid model for radiation to yield a general relativistic radiation magnetohydrodynamic (GRRMHD) model in axisymmetry (\citealt{Sadowski+2017}) and 3D (\citealt{SadowskiGaspari2017}). Simulations have generally confirmed the picture of a RIAF perturbed by radiative cooling, although details of the transition to radiatively efficient thin disks are still uncertain.

We have developed a numerical method, \bhlight{}, for solving the GRRMHD equations with a Monte Carlo method to provide a direct solution to the frequency-dependent radiative transport equation, including emission, absorption, and Compton scattering (\citealt{Ryan+2015}). We introduced \ebhlight{} to include the electron heating scheme of \cite{Ressler+2015} with the Coulomb coupling in \cite{Ryan+2017}. Surveying $\dot{m}$ for $M=10^8M_{\odot}$, \cite{Ryan+2017} found radiative cooling to be significant for $\dot{m} \gtrsim 10^{-5}$, with high-energy spectra progressively hardening and previously distinct Compton bumps merging to form a smooth power-law tail with increasing $\dot{m}$. \ebhlight{} allows us to model optically thin RIAFs in axisymmetry without substantial approximation to the radiation physics, although our model remains sensitive to the electron thermodynamics, and to our assumption that the electron distribution function is thermal.

% This work
In this work we study a suite of global axisymmetric GRRMHD \ebhlight{} simulations to interpret time-averaged spectral and imaging observations of M87. In Section \ref{sec:eqns} we describe the governing equations, and in Section \ref{sec:numerics} we describe our numerical implementation and present a test of our code. Section \ref{sec:results} presents our models and results. Section \ref{sec:discussion} discusses these results in the context of current and upcoming observations, and Section \ref{sec:conclusion} concludes.

\section{Governing Equations}
\label{sec:eqns}

\bhlight{} solves the equations of GRRMHD in stationary spacetimes with frequency-dependent radiative transport. Electron temperatures are evolved self-consistently according to a plasma-dependent heating prescription. Photon-electron emission, absorption, and scattering couple the matter and radiation. In this section, we adopt units such that $GM = c = 1$. Throughout this work, we express lengths in units of $r_{\rm G} \equiv GM/c^2$ and times in units of $t_{\rm G} \equiv GM/c^3$.

\subsection{MHD}

The equations of GRMHD for conservation of mass, energy-momentum, and magnetic flux along with the no-monopoles constraint (e.g.\ \citealt{Gammie+2003}) take the forms, respectively,

\begin{align}
\partial_t \left( \sqrt{-g} \rho_0 u^t \right) &= -\partial_i \left( \sqrt{-g} \rho_0 u^i \right), \label{eqn:massConservation} \\
\begin{split}\label{eqn:stressEnergyConservation}
    \partial_t \left( \sqrt{-g} T^t_{~\nu} \right) &={} -\partial_i \left( \sqrt{-g} T^i_{~\nu} \right) + \sqrt{-g}T^{\kappa}_{~\lambda} \Gamma^{\lambda}_{~\nu\kappa}\\
& \quad - \sqrt{-g}R^{\mu}_{~\nu;\mu},
\end{split}\\
\partial_t \left( \sqrt{-g} B^i \right) &= \partial_j \left[ \sqrt{-g} \left( b^j u^i - b^i u^j \right) \right], \label{eqn:fluxConservation} \\
\partial_i \left( \sqrt{-g} B^i \right) &= 0. \label{eqn:monopoleConstraint}
\end{align}
with $\rho_0$ the rest-mass density, $u^{\mu}$ the fluid four-velocity, $\Gamma^{\mu}_{~\nu \lambda}$ the Christoffel symbols, the total fluid stress-energy tensor
\begin{align}
T^{\mu}_{~\nu} &= \left( \rho_0 + u + P + b^{\lambda}b_{\lambda}\right)u^{\mu}u_{\nu} \\
&+ \left(P + \frac{b^{\lambda}b_{\lambda}}{2} \right)g^{\mu}_{~\nu} - b^{\mu}b_{\nu}, \nonumber
\end{align}
$b^{\mu}$ the magnetic field four-vector, $u$ the total fluid internal energy density, $P = (\gamma-1)u$ the total fluid pressure, and the radiation stress-energy tensor
\begin{align}
R^{\mu}_{~\nu} = \int \frac{d^3 p}{\sqrt{-g}p^t}p^{\mu}p_{\nu} \left(\frac{I_{\nu}}{h^4\nu^3}\right),
\end{align}
($p^{\mu} \equiv$ photon four-momentum, $I_{\nu} \equiv $ specific intensity) the four-divergence of which ($G_{\nu} \equiv R^{\mu}_{~\nu;\mu}$) gives the radiation four-force applied to the total fluid.

\subsection{Two-temperature Thermodynamics}

To obtain both proton and electron temperatures, we solve for the electron entropy in addition to the total fluid energy as in \cite{Ressler+2015}. The first law of thermodynamics for the electrons in a coordinate basis is
\begin{align}
\frac{\rho^{\gamma_e}}{\gamma_e - 1} u^{\mu} \partial_{\mu} \kappa_e &= f_e Q_H + Q_C - u^{\nu}R^{\mu}_{~\nu;\mu}, \label{eqn:electronEntropy}
\end{align}
%where $\kappa_e \equiv \exp((\gamma_e-1)s_e) = P_e/\rho_0^{\gamma_e}$ ($s_e \equiv$ electron entropy), $Q_H$ and $Q_C$ are, respectively, dissipative (viscous/resistive) and Coulomb \citep{StepneyGuilbert1983} volumetric heating rates. The last term represents exchange of energy between photons and electrons.
where $\kappa_e \equiv \exp((\gamma_e-1)s_e) = P_e/\rho_0^{\gamma_e}$ ($s_e \equiv$ electron entropy), $Q_H$ and $Q_C$ are, respectively, dissipative and Coulomb \citep{StepneyGuilbert1983} volumetric heating rates. The factor $f_e$ is a function of local plasma properties and represents the fraction of total dissipation that is applied to electrons by the assumed dissipation mechanism. Throughout this work we adopt the $f_e$ of Howes 2010, which attributes dissipation to kinetic damping at small scales. This $f_e$ leads to dissipation being captured mostly by ions at high plasma $\beta$, and mostly by electrons at low plasma $\beta$. Observables are generally sensitive to $f_e$; see \cite{Chael+2018} for a study of the effects of varying the $f_e$ prescription in GRMHD models of Sgr A*. The last term in Equation \ref{eqn:electronEntropy} represents the exchange of energy between photons and electrons. Once $u$ (Equation \ref{eqn:stressEnergyConservation}) and $u_e$ (Equation \ref{eqn:electronEntropy}) are known, the proton internal energy is given by $u_p = u - u_e$; $u_p$ is not evolved separately. For our numerical implementation of Equation \ref{eqn:electronEntropy}, see Section \ref{sec:electronNumerics}.

Typically in hot accretion disks, the electrons are relativistic while the protons are nonrelativistic; the two species thus have different adiabatic indices. We adopt three constant values: $\gamma_p = 5/3$ for protons, $\gamma_e = 4/3$ for electrons, and $\gamma = 13/9$ for the total fluid. This is probably reasonably accurate in the inner region of hot accretion flows; see \citealt{Sadowski+2017} for results of two-temperature electron heating employing a more sophisticated treatment of adiabatic indices. %The proton internal energy density (needed for the Coulomb heating rate) is $u_p = u - u_e$.

\subsection{Covariant Radiation Transport}

We solve the radiative transfer equation in invariant form
\begin{align}
\frac{d}{d\lambda}\left(\frac{I_{\nu}}{\nu^3}\right) &= \frac{\eta_{\nu}}{\nu^2} -(\nu \chi_{\nu})\frac{I_{\nu}}{\nu^3}, \label{eqn:radiativeTransfer}
\end{align}
along geodesics described by
\begin{align}
\frac{dx^{\mu}}{d\lambda} &= k^{\mu}, \label{eqn:photonMotion} \\
\frac{dk^{\lambda}}{d\lambda} &= -\Gamma^{\lambda}_{~\mu\nu}k^{\mu}k^{\nu}, \label{eqn:geodesic}
\end{align}
where $x^{\mu}$ is the spacetime coordinate of a photon, $k^{\mu}$ the corresponding wavevector, and $d \lambda$ the affine parameter along the geodesic. $\eta_{\nu}$ and $\chi_{\nu}$ are the emission and absorption coefficients, respectively, and include contributions from scattering. We consider thermal synchrotron emission and absorption, and Compton scattering.

\section{Numerical Method}
\label{sec:numerics}

\ebhlight{} (\citealt{Ryan+2017}) solves the equations of GRRMHD with frequency-dependent radiation transport. Here we provide an overview of the numerical implementation, emphasizing the interplay between the electron thermodynamics \citep{Ressler+2015} and the radiation. The one-temperature GRRMHD method \bhlight{} is described in detail in \cite{Ryan+2015}. 

\subsection{GRMHD}

\ebhlight{}'s fluid sector is based on the GRMHD scheme \harm{} \citep{Gammie+2003}. \harm{} is a relativistic second-order explicit shock-capturing scheme for stationary spacetimes. Magnetic monopoles are suppressed to roundoff error through flux-interpolated constrained transport \citep{Toth2000}.

The radiation four-force, the cumulative representation of emission, absorption, and scattering (equivalently, the divergence of the radiation stress-energy tensor), is applied in a first-order operator-split fashion to the total energy and momentum of the MHD sector. While this is inferior to the second-order accuracy elsewhere in the GRMHD sector, \ebhlight{} is designed for RIAF problems for which the radiative cooling time is long compared to the timestep.

\ebhlight{} uses an axisymmetric implementation of \harm{}. Accretion disks around black holes are nearly axisymmetric on large scales, at least in the absence of disk tilts and strong vertical magnetic flux. However, the magnetorotational turbulence responsible for accretion in our simulations is changed qualitatively by axisymmetry. In particular, at least two pathologies appear: (1) in the absence of net fields, turbulence decays due to the antidynamo theorem (2) power in 2D MHD turbulence cascades toward larger scales. In practice these effects limit our simulation runtime. \cite{GuanGammie2008} studied magnetorotational turbulence in an axisymmetric local model. They found that turbulence decays on a timescale $t_{\rm D} \sim 20 \Omega^{-1}$ (note that $t_{\rm D} \sim 600 t_{\rm G}$ at $r = 10 r_{\rm G}$) and increases in amplitude with numerical resolution. In addition, axisymmetry washes out nonaxisymmetric structures, which have been shown to give rise to observable variability in \cite{Dolence+2012}.

However, the resolution dependence identified by \cite{GuanGammie2008} can be used to tune the stress in axisymmetric simulations. Existing global simulations suggest that a good resolution to recover 3D stresses in axisymmetric models occurs between $256^2$ and $512^2$ zones with a \harm{}-like mesh refinement. For example, compare the accretion rate of the $256^2$ axisymmetric fiducial model of \cite{McKinneyGammie2004} to the accretion rate of the $192\times192\times128$ disk simulation in \cite{Shiokawa+2012}; over the common time domain, the accretion rates are within a factor of a few. Additionally, early work modeling Sagittarius A* with GRMHD simulations (\citealt{Moscibrodzka+2009}) was performed in axisymmetry; differences with subsequent 3D simulations may be dominated by choice of electron temperature prescription rather than internal stresses (see e.g. \citealt{Moscibrodzka+2013} for a direct comparison to \citealt{Moscibrodzka+2009}). 

%To summarize, history and evidence suggest that axisymmetric global simulations can be profitably applied to studying disk properties and observables not sensitive to non-axisymmetric structures. 
Here we conservatively consider spectra, millimeter image sizes, and jet powers. Observables such as variability and more detailed imaging, along with alternative disk models incorporating tilt and strong polodial magnetic fields, are best left to 3D simulations. Finally, in an attempt to further suppress errors introduced by axisymmetric fluid evolution, we initialize our simulations from axisymmetrized final states of 3D GRMHD simulations (see Section \ref{subsec:initialconditions} for details).

% Radiation four-force

\subsection{Radiation}

The radiation field is discretized into Monte Carlo samples, hereafter ``superphotons,'' based on the relativistic radiative transfer scheme \grmonty{} \citep{Dolence+2009}. Each superphoton possesses the usual properties of a photon (position $x^{\mu}$, wavevector $k^{\mu}$) along with a weight $w$ corresponding to the number of constituent photons. In contrast to \grmonty{}, each superphoton is emitted with equal total energy, i.e. $h \nu w = \mathrm{const}$ (\citealt{AbbottLucy1985}), which tends to provide the highest accuracy at fixed computational expense for Monte Carlo radiation hydrodynamics. The radiation boundary is typically further in than the fluid boundary. This allows us to place the outer boundary of GRMHD evolution far from the black hole to avoid spurious fluid boundary effects while avoiding the computational expense of integrating large numbers of superphotons along nearly straight geodesics through regions with negligible radiation-matter interactions.

% Geodesic

\ebhlight{} integrates superphotons along geodesics using a second-order explicit step on the fluid timestep. Although our second-order scheme requires only one evaluation of the Christoffel symbols per geodesic update, in practice this is the dominant computational cost in accretion disk simulations. 

% Emission

Superphotons are created by sampling the total emissivity of the plasma in the fluid frame over frequency and angle. \ebhlight{} is physics-agnostic in this regard. While we have implemented the thermal electron synchrotron emissivity of \cite{Leung+2011}, new emissivities are readily introduced.

% Absorption and scattering

Absorption and scattering are incorporated probabilistically. Following integration along geodesics, optical depths to absorption and scattering are calculated based on the traversed affine parameter $\Delta \lambda$. These optical depths are sampled to determine if an interaction has taken place, and if so whether it was absorption or scattering. If so, the superphoton is pushed back along its geodesic to the site of the interaction and the interaction is processed. For absorption, the superphoton is completely absorbed. For scattering, a bias parameter is used such that superphotons are scattered more frequently, but only with a fraction of their weight $w$. This process generates an additional superphoton for each biased scattering, and greatly enhances resolution in the radiation field when the optical depth to scattering is small but the amplification factor due to scattering is large.

The desired number of superphotons per MPI process is specified as a runtime parameter in \ebhlight{}. Superphoton resolution (i.e.\ weight $w$) is controlled dynamically in two ways so as to recover this desired number of superphotons in a time-averaged sense. First, the energy per superphoton is adjusted to control the emitted number of superphotons, based on the difference in emitted and absorbed superphotons relative to the light crossing time of the radiation region. Second, the scattering bias is adjusted such that each emitted superphoton scatters approximately once. With less frequent scattering, resolution is lost at higher photon frequencies. With more frequent scattering, the numerics go critical, analogous to a fission reactor meltdown.

The usual signature of insufficient superphoton resolution for our RIAF models is ``supercooling,'' in which superphoton weights are too large to accurately sample emission and scattering; there are too few interactions per cooling time. Zones may then be cooled to negative internal energies, which after GRMHD fixup routines results in energy nonconservation, or at least spurious electron heating. We monitor supercooling to ensure that this anomalous energy is not a significant part of the radiation energy budget, i.e.\ that we have a sufficient number of superphotons per MPI process and a sufficient number of MPI processes. Supercooling could also result from local cooling timescales shorter than the global simulation timestep. Although we do not encounter such short cooling timescales in our target application, implicit Monte Carlo methods (e.g. \citealt{RothKasen2015}) have been developed to address this issue.

\subsection{Electron Thermodynamics}
\label{sec:electronNumerics}

We model the electron temperature with the self-consistent electron heating scheme of \cite{Ressler+2015}. Dissipative heating in ideal MHD schemes such as \harm{} is present as grid-scale truncation error. Here we evaluate that heating by advecting the total entropy of the fluid simultaneously with the traditional \harm{} update of conserved mass, energy, and momentum. After the step, the advected and evolved total entropies are compared; this difference is the heating rate. A fraction of this heating, evaluated based on the local plasma prescription of \cite{Howes2010}, is then applied the electron entropy, which is itself advected and which provides the electron temperature. Rather than depositing the remaining fraction of heating in a proton internal energy variable, we simply evaluate the proton energy as the total internal energy minus the electron internal energy.

In \ebhlight{}, the electron entropy variable provides the electron temperature used in evaluating emissivities, absorptivities, and scattering events. In addition, the radiation exchanges energy directly with this electron entropy, rather than with the total fluid internal energy. We apply the timelike component of the radiation four-force in the comoving frame, $u^{\mu}G_{\mu}$, to the electron entropy in a similar first-order operator split fashion as we apply $G_{\mu}$ to the total fluid stress-energy tensor. 

We now also include Coulomb heating as a second-order operator-split explicit update to the electron entropy. Evaluating electron and proton temperatures from the total fluid internal energy and the electron entropy, we calculate the Coulomb heating rate of \cite{StepneyGuilbert1983}. We then update the electron entropy in accordance with this heating; the total internal energy is unchanged.

We now summarize the entire heating process over a timestep. First, total dissipation is captured by differencing the total entropy evaluated at the initial state and advected to the final state, and the total entropy evaluated from the final state. The \cite{Howes2010} prescription determines what fraction of this dissipation is applied to the electron entropy variable. The updated proton internal energy, which is only needed for evaluating the Coulomb interaction rate, is then known from the updated total and electron internal energies. Finally, energy is added to or subtracted from the electron entropy variable according to the Coulomb interaction rate, without changing the total internal energy.

\harm{}-like codes require fixup routines, which enforce minimum densities of rest mass and energy, to avoid instability during the fluid integration. Similar fixup routines are used for the electron entropy by \cite{Ressler+2015}. Our inclusion of Coulomb heating, which requires positive proton and electron temperatures, motivates somewhat different fixup routines than those in \cite{Ressler+2015}. In particular, we enforce the ratio of proton-to-electron temperature $(T_p/T_e)_{\mathrm{min}}>0.01$. In the radiation sector, we enforce $\Theta_e \equiv k_B T_e / m_e c^2 < 1000$ to avoid failures in sampling Compton scattering. Additionally, we forbid radiation interactions in the highly magnetized funnel region ($b^2/\rho > 1$), where \harm{}-like total energy codes cannot accurately represent even total fluid thermodynamics.\footnote{At least within the GRMHD model, densities in the funnel region seem to have no lower limit; material either falls onto the black hole or is ejected to infinity, and the funnel wall itself is apparently stable to long-wavelength instabilities, at least for steady dipolar fields and an ideal fluid (\citealt{McKinneyBlandford2009}). As a result, densities in the funnel are usually set by the numerical floor required for stability. Without additional physics such as pair production (e.g.\ \citealt{Moscibrodzka+2011}), our model constrains us to suppose that densities in the funnel are too low to lead to significant radiation.}

\subsection{Numerical Verification: $e$-$p$-$\gamma$ Thermalization}

We now present a test verifying our implementation, emphasizing Coulomb coupling and electron-photon interactions. Additional tests of the electron heating and radiation MHD sectors may be found in \cite{Ressler+2015} and \cite{Ryan+2015}. We consider a one-zone model for the thermalization of electrons, protons, and photons. Electrons and protons interact through Coulomb collisions, while (only in this test) electrons and photons interact through a bremsstrahlung-like emissivity,
\begin{align}
j_{\nu} = N n_e n_p T_e^{-1/2} \exp{\left(\frac{-h \nu}{k_B T_e}\right)},
\end{align}
where $n_e = n_p$ are the electron and ion number densities and $N = 5.4\times10^{-39}~\mathrm{cm^{3}~K^{1/2}~s^{-1}~Sr^{-1}~Hz^{-1}}$. Thermal absorption is included. This test is similar to the thermalization problem in \cite{Sadowski+2017}, except that we consider the full multifrequency problem through a frequency-dependent opacity.

We set the mass density $\rho = 2\times10^{-4}~\mathrm{g~cm^{-3}}$ and initial proton and electron temperatures $T_{p,0} = 10^{8}~\mathrm{K}$ and $T_{e,0} = 10^{7}~\mathrm{K}$. No radiation is present initially. We set the Coulomb logarithm $\log \Lambda = 0.01$ (present in $Q_{\mathrm{coul}}$) to enforce comparable Coulomb and emission timescales for an equilibrium temperature at which the radiation pressure does not overwhelm the gas pressure. We set $\gamma = 13/9$, $\gamma_e = 4/3$, and $\gamma_p = 5/3$. 

\begin{figure}
\centering
\includegraphics[width=\figwidth]{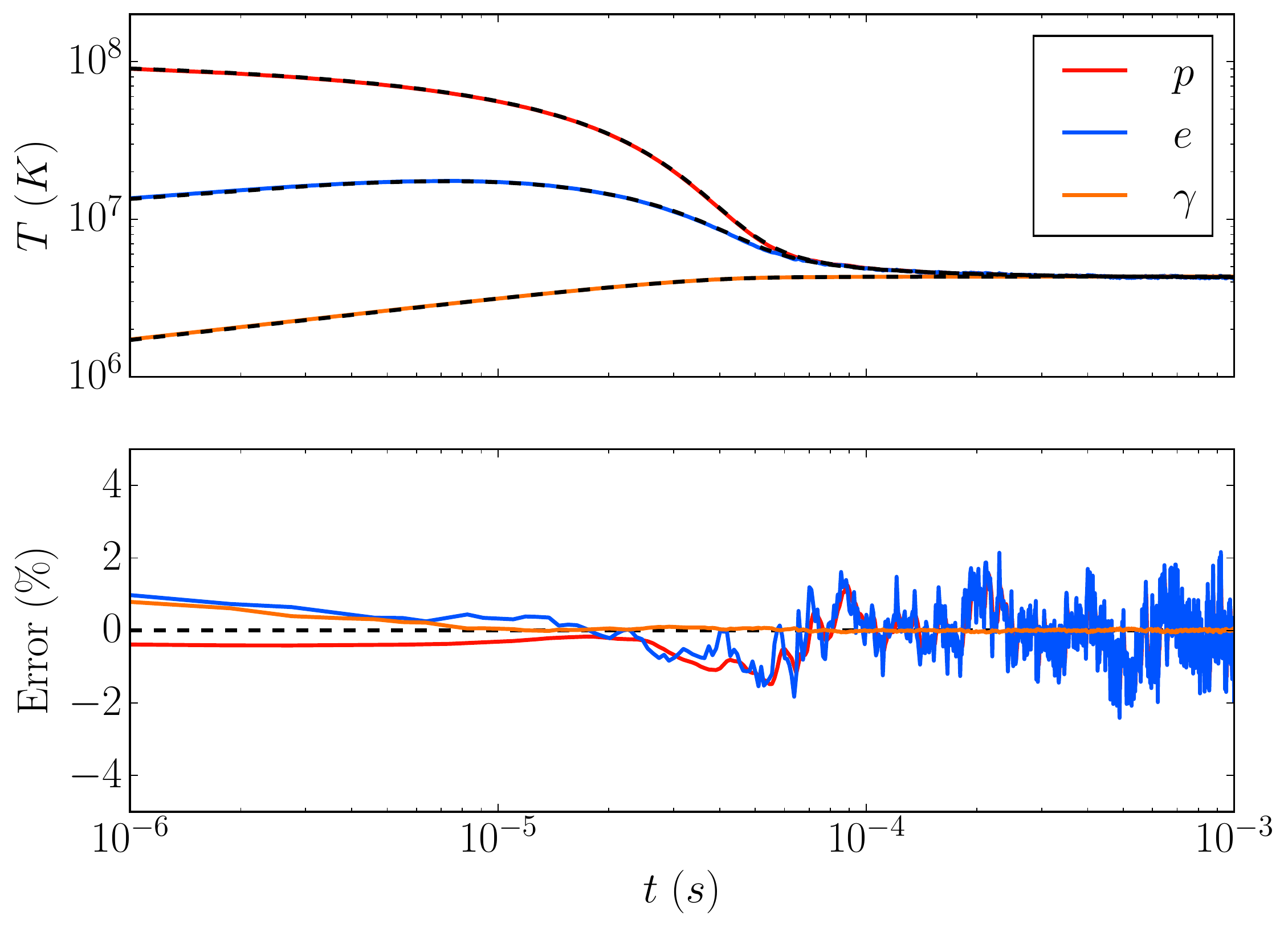}
\caption{Proton, electron, and radiation temperatures for the thermalization test, along with relative errors. Solid lines denote the \ebhlight{} solution, while dashed lines give the semianalytic solution. Relative error is small, and at late time is dominated by Monte Carlo noise.}
\label{fig:thermalization_test}
\end{figure}

We construct a semianalytic solution by solving the integro-differential equations
\begin{align}
%a &= b
\frac{d T_e}{d t} &= \frac{\gamma_e-1}{n_e k_B}\left( Q_{\mathrm{coul}} - \int  d \nu \frac{d u_{\nu}}{d t}\right), \\
\frac{d T_p}{d t} &= -\frac{\gamma_p-1}{n_p k_B}Q_{\mathrm{coul}}, \\
\frac{d u_{\nu}}{d t} &= 4 \pi j_{\nu}(T_e) \left(1 - \frac{c u_{\nu}}{4 \pi B_{\nu}(T_e)} \right),
\end{align}
where the specific radiation energy density $u_{\nu}$ is discretized over frequency. We compare numerical output with this semianalytic solution in Figure \ref{fig:thermalization_test}, and find good agreement.

\section{Results}
\label{sec:results}

We set out to model M87 for several parameter choices in order to identify which, if any, are consistent with observations. Our simulations are the four combinations of black hole masses $M/M_{\odot} = (3.3\times10^9, 6.2\times10^9)$ and spins $a_{\star} = (0.5, 0.9375)$. We assume a distance $D=16.7$ Mpc (\citealt{Blakeslee+2009}). The accretion rate is iterated until we recover the 230 GHz flux measured for M87 from \cite{Doeleman+2012}, who found $F_{\nu,230~\mathrm{GHz}} = 0.98 \pm 0.04$ Jy. Of all the observed frequencies, this is probably the best choice for normalizing our models. The source is probably optically thin at 230 GHz; 230 GHz synchrotron emission is dominated by relativistic electrons near the black hole (as opposed to IR and X-ray emission, which may have contributions from outside our simulation domain), and \cite{Doeleman+2012} demonstrated that at least some of this emission originates in a from a compact region at $r \lesssim 10 r_{\rm G}$. The observer inclination angle is fixed at $\theta = 20\degree$ (e.g.\ \citealt{HeinzBegelman1997})\footnote{\citealt{Dexter+2012} found, for millimeter images of M87 derived from GRMHD simulations, that decreasing $\theta$ caused images to be more ring-shaped and less Gaussian. Image size was less sensitive to $\theta$.}.

\subsection{Initial conditions}
\label{subsec:initialconditions}

We use the horizon-penetrating Modified Kerr-Schild coordinates of \cite{McKinneyGammie2004}, with $\theta$-refinement parameter $h=0.3$. We adopt a spatial resolution of $388\times256$ zones.  We target $\sim 10^7$ superphotons at saturation. The inner boundary is placed such that five zones (one fluid reconstruction stencil) are inside the event horizon, while the outer fluid boundary is placed at $200 r_{\rm G}$. The outer radiation boundary is set to $40 r_{\rm G}$ for $M/M_{\odot} = 6.2\times10^9$ and $100 Gr_{\rm G}$ for $M/M_{\odot} = 3.3\times10^9$, beyond which radiative interactions are negligible in our model (there are radiative contributions at larger radii for $M/M_{\odot} = 3.3\times10^9$ simulations; see Table \ref{tab:models}). The larger outer radiation boundary in $M/M_{\odot} = 3.3\times10^9$ contributes to the decreased signal-to-noise in Figure \ref{fig:nuFnu}.

We initialize our torii using axisymmetrized 3D data from two 3D two-temperature GRMHD simulations ($a_{\star} = 0.5$ and $a_{\star} = 0.9375$, otherwise similar) run for $10,000~t_{\rm G}$. This allows the electron temperatures to saturate due to dissipative heating through a larger region of the disk ($r \lesssim 12 r_{\rm G}$) than is possible with the limited runtimes available in axisymmetric models. Fluid mass, internal energy, and velocity, along with electron entropy, are averaged in $\phi$, while the magnetic field is first converted to a magnetic vector potential, which is averaged in $\phi$ and then differentiated to recover an axisymmetric divergence-free magnetic field. No radiation is present initially, and the radiation field and superphoton number density equilibrate on the light crossing time. Each simulation is run for $1000~t_{\rm G}$; time averages are begun at $600~t_{\rm G}$, by which time bolometric luminosity $L$ and radiative efficiency $\epsilon$ are relatively steady (see Figure \ref{fig:diags}). 

To first approximation, the saturated state of magnetorotational turbulence is determined by one parameter, the net vertical field strength (e.g.\ \citealt{Salvesen+2016a, Salvesen+2016b}). The magnetic field configuration in our models is set by the poloidal flux in the initial conditions of the 3D simulations. In our models this flux is relatively weak (that is, we are simulating a SANE rather than a Magnetically Arrested Disk (MAD) flow; see e.g. \citealt{Narayan+2012} for a comparison). As a consequence, the polar jets are relatively weak as well; see Section \ref{sec:netflux} for a discussion of the effects of stronger net vertical fields on observable quantities. 

%These models have relatively weak polar jets; see \ref{sec:netflux} for a discussion of how stronger net vertical fields might affect our models.  

%The accretion rate for which the time-averaged 230 GHz flux agrees with the EHT result is determined by rootfinding. For each mass and spin, \ebhlight{} simulations are performed in sequence at different $\dot{m}$ until the simulation and EHT $F_{\nu,230\mathrm{GHz}}$ agree to within $5 \%$, roughly the error reported by \cite{Doeleman+2012}. This fully specifies the accretion disk for each mass and spin.
We set the accretion rate from the scale-free GRMHD evolution by fixing the black hole mass and then the mass of the accretion disk. The accretion rate for which the time-averaged 230 GHz flux agrees with the EHT result is determined by root finding. For each mass and spin, \ebhlight{} simulations are performed in sequence at different $\dot{m}$ until the simulation and EHT $F_{\nu,230~\mathrm{GHz}}$ agree to within $5 \%$, roughly the error reported by \cite{Doeleman+2012}. This fully specifies the accretion disk for each mass and spin.

\subsection{Diagnostics}

We employ weighted shell averages $\langle f \rangle_w$ such that
\begin{align}
\langle f \rangle_w = \frac{\int fw\sqrt{-g} dx^2 dx^3}{\int w\sqrt{-g} dx^2 dx^3}.
\end{align}
We use similar notation to denote unweighted averages inside a maximum radius $r_{\mathrm{out}}$:
\begin{align}
\langle f \rangle_{r_{\mathrm{out}}} = \frac{\int^{r_{\mathrm{out}}} f \sqrt{-g} dx^2 dx^3}
{\int^{r_{\mathrm{out}}} \sqrt{-g} dx^2 dx^3};
\end{align}
see the appendix in \citealt{Farris+2010} for a discussion of the transformation properties of integrals over $\sqrt{-g} dx^2 dx^3$. For quantities representing ratios ($T_p/T_e$, $\beta$, $Q_{\mathrm{coul}}/Q_{\mathrm{visc}}$, $t_* \equiv u_*/Q_*$), $\langle A/B \rangle$ implies $\langle A \rangle/\langle B \rangle$. We do this so that isolated, small values of the denominators do not overly bias the averages.

We define a disk aspect ratio $H/R = \tan \theta_d$, where we follow \cite{McKinney+2012} and define
\begin{align}
\theta_d^2 \equiv \langle (\theta - \theta_0)^2 \rangle_{\rho},
\end{align}
with $\theta_0 =  \langle\theta \rangle_{\rho}$. 
Additionally, we define an accretion rate
\begin{align}
\dot{M} =- \int \rho u^1 \sqrt{-g} dx^2 dx^3,
\end{align}
and a bolometric luminosity
\begin{align}
L = - \int R^1_{~0} \sqrt{-g} dx^2 dx^3,
\end{align}
with a corresponding radiative efficiency $\epsilon \equiv L/\dot{M}$. Here, $L$ and $\dot{M}$ are evaluated at the outer radiation and inner fluid radial boundaries, respectively, and $\epsilon$ therefore contains a delay corresponding to the light crossing time and ignores the effects of outflows. $R_{95} \equiv$ the radius inside of which $95\%$ of the bolometric luminosity is generated. 

We compute jet power by integrating internal energy and electromagnetic fluxes in the jet region where $b^2/\rho > 1$,
\begin{align}
P_{\rm J} &= \int{\left[ \left(\gamma u + b^2\right)u^1u_0 -b^1b_0\right]\sqrt{-g} dx^2 dx^3}
\end{align} 
This procedure omits any fluxes outside the Poynting jet (e.g.\ \citealt{BlandfordPayne1982}). This integral is performed over spherical shells and averaged over $r \in [20, 40]$. Averaged quantities for our simulations are summarized in Table \ref{tab:models}.

\begin{table*}[ht]
\centering
\begin{tabular}{ccccccccc}%{p{0.25\linewidth}p{0.25\linewidth}p{0.25\linewidth}}
\hline
% Mdot 
Label & $\dot{m}$ & $\epsilon$ & $\langle {\Theta_e} \rangle_J$ & ${L_{\mathrm{em}}}/{L_{\mathrm{sc}}}$ & $\langle Q_{\mathrm{coul}} \rangle_{10}/\langle Q_{\mathrm{visc}} \rangle_{10}$ & $\langle H/R \rangle_{10}$ & $R_{95}~(r_{\rm G})$ & $P_{\mathrm{J}}~(\mathrm{erg~s^{-1}})$ \\
\hline
{\tt M3a05} & $2.2\times10^{-5}$ & $1.6\times10^{-2}$ & 5.1 & 0.33 & 0.03 & 0.26 & 67 & $2.3\times10^{40}$\\ % 30 %
%PJM = 6.424955e+36 erg s^-1
{\tt M3a09} & $8.2\times10^{-6}$ & $2.4\times10^{-2}$ & 8.7 & 0.75 & 0.04 & 0.32 & 88 & $5.0\times10^{41}$\\ % 20%
{\tt M6a05} & $9.2\times10^{-6}$ & $6.7\times10^{-3}$ & 9.3 & 1.4 & 0.013 & 0.26 & 31 & $1.6\times10^{40}$\\ % 36%
% PFM = 7.7e36
{\tt M6a09} & $5.2\times10^{-6}$ & $1.2\times10^{-2}$ & 14 & 1.5 & 0.024 & 0.32 & 12 & $5.1\times10^{41}$\\ % 18%
% PJM = 1.0e38
\hline
\end{tabular}
\caption{For each simulation, time-averaged fluid and radiation quantities: accretion rate, radiative efficiency, emissivity-weighted electron temperature, ratio of emitted and scattered photon contributions to bolometric luminosity (roughly the inverse of Compton $y$), ratio of Coulomb to dissipative heating inside $r = 10 r_{\rm G}$, disk thickness averaged inside $r = 10 r_{\rm G}$,  radius of region contributing to luminosity, and jet power.}
\label{tab:models}
\end{table*}

\subsection{Intrinsic Model Properties}

Our models, at $\dot{m} \sim 10^{-5}$, occupy an interesting range of accretion rates, and at black hole masses somewhat higher than the low-spin ($a_{\star} = 0.5$), $M/M_{\odot} = 10^8$ black hole considered in \cite{Ryan+2017}. Therefore, in this section we elaborate on the thermodynamic state of the accretion disk in our models. Note that care should be taken in extracting trends with mass and spin from this simulation suite, as the accretion rate for each simulation is set by the 230 GHz flux.

\begin{figure}
\centering
\includegraphics[width=\figwidth]{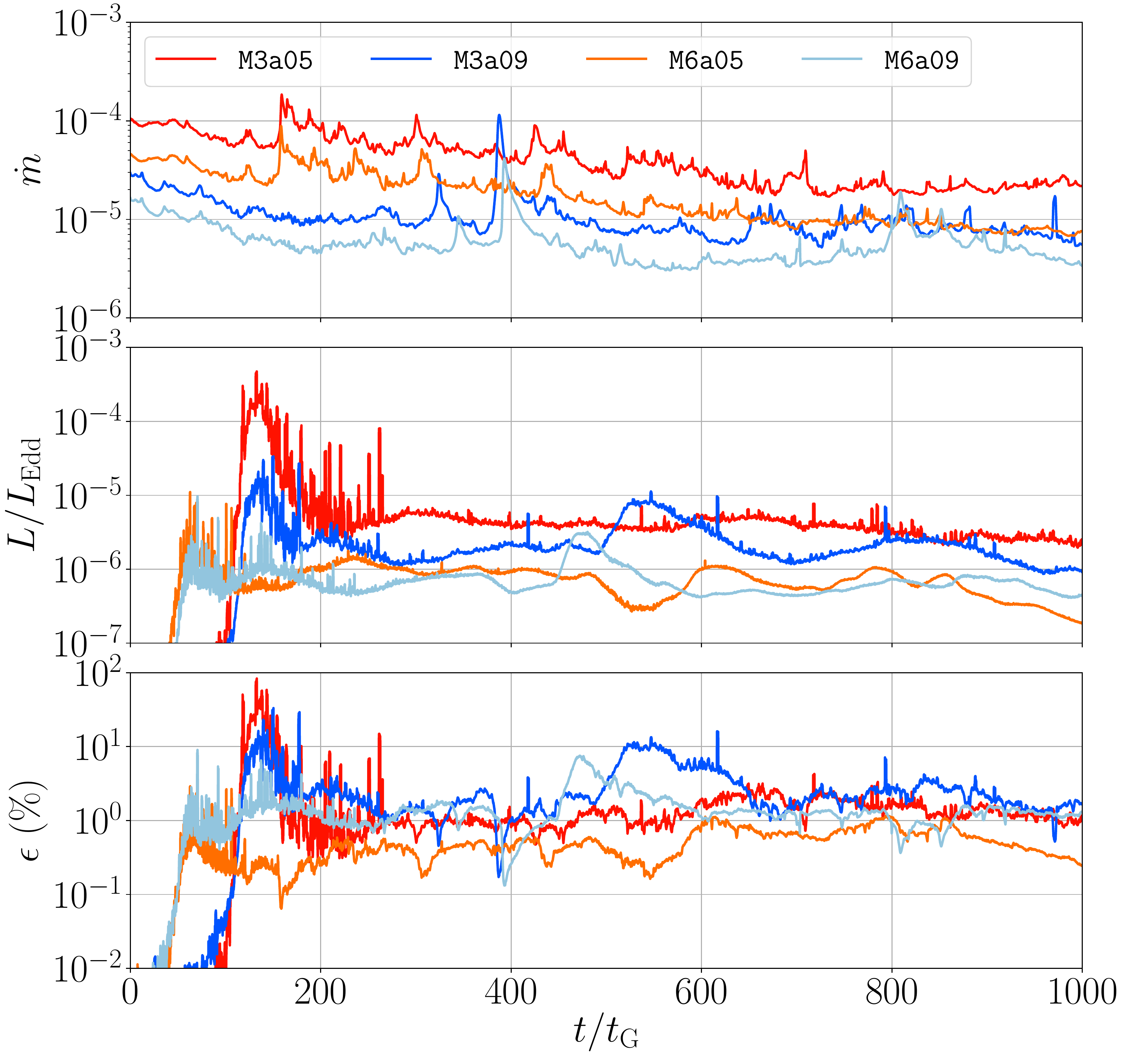}
\caption{Accretion rate, luminosity, and radiative efficiency as a function of time for all models.}
\label{fig:diags}
\end{figure}

Accretion rate (in Eddington units), luminosity, and radiative efficiency are shown as a function of time for all models in Figure \ref{fig:diags}. For the black hole masses and spins considered here, $\dot{m}$ varies by a factor $\sim 20$ across models. Radiative efficiency, on the other hand, varies by $\lesssim 4$. While axisymmetry probably enhances the variability of GRMHD simulations, there is no dramatic secular trend in \mdot{}, $L$, or $\epsilon$ over our time integration window ($600/t_{\rm G} < t < 1000 t_{\rm G}$).

\begin{figure}
\centering
\includegraphics[width=\figwidth]{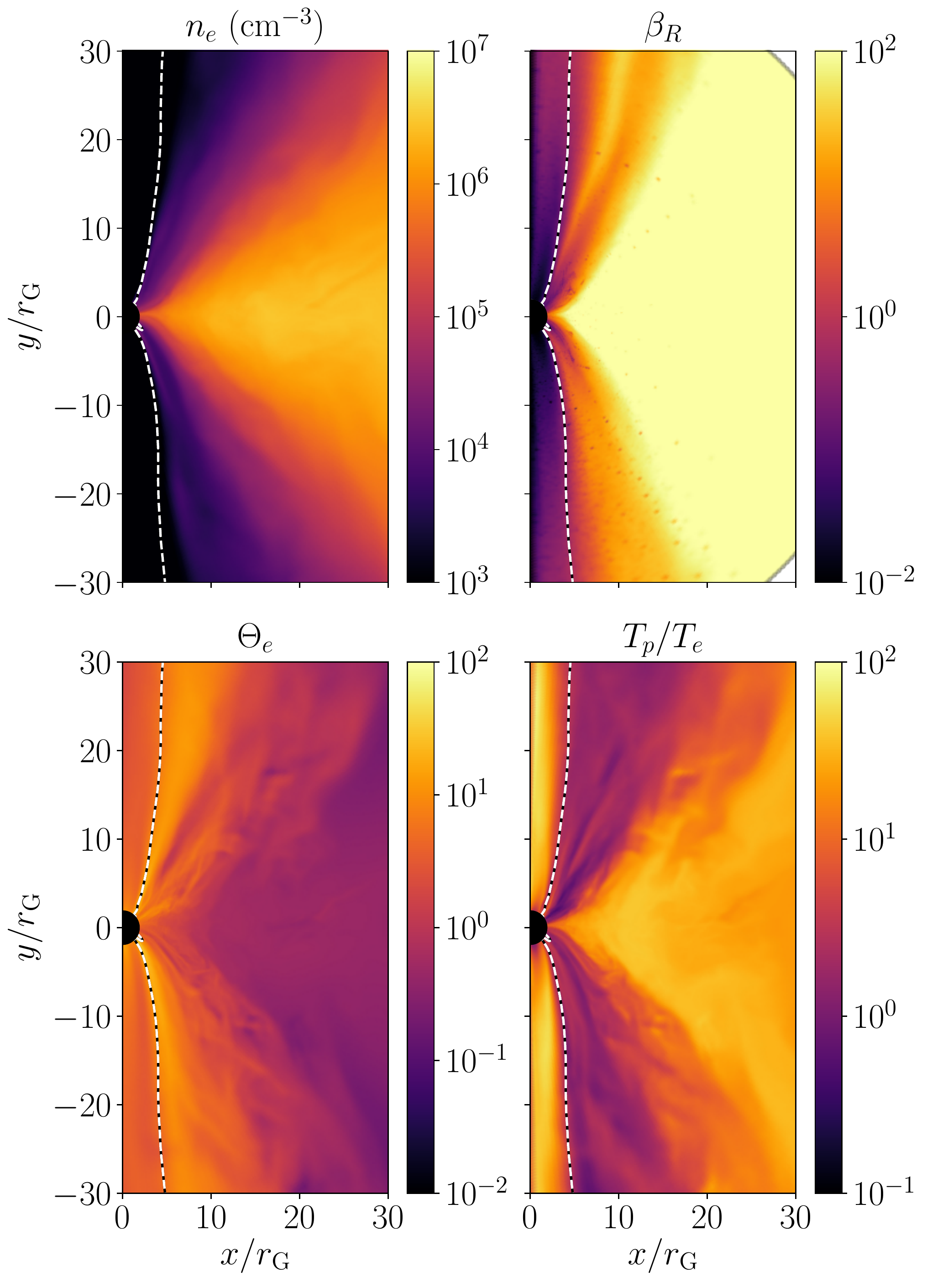}
\caption{Azimuthal slices of time-averaged density, radiation to gas pressure ratio $\beta_R \equiv P_g/R^{(0)(0)}$ where $R^{(0)(0)}$ is evaluated in the comoving frame and $P_g$ is the total gas pressure, electron temperature, and proton-to-electron temperature ratio for simulation {\tt M6a05}. Dashed lines denote the boundary of the magnetized funnel, where $b^2/\rho > 1$.}
\label{fig:avg_grmhd}
\end{figure}

Time-averaged density, radiation to gas pressure ratio $\beta_R \equiv R^{00}/(3 P_g)$ where $R^{00}$ is evaluated in the fluid frame, dimensionless electron temperature $\Theta_e$, and proton-to-electron temperature ratio $T_p/T_e$ from ${\tt M6a05}$ are shown in Figure \ref{fig:avg_grmhd}. Despite cooling, for this accretion rate and runtime the ions in the torus continue to resemble a nonradiative hot, geometrically thick accretion flow. Consistent with the pure GRMHD calculations of the self-consistent electron heating of \cite{Ressler+2015}, midplane electrons are generally cold while coronal electrons are hot. Notice however that the inclusion of Coulomb heating in this model enhances electron temperatures preferentially in the midplane, where collision times are shorter. The variable two-temperature nature of the flow is shown explicitly in the spatial dependence of $T_p/T_e$, which is approximately unity at the funnel wall and $\sim 10-30$ in the midplane.

\begin{figure}
\centering
\includegraphics[width=\figwidth]{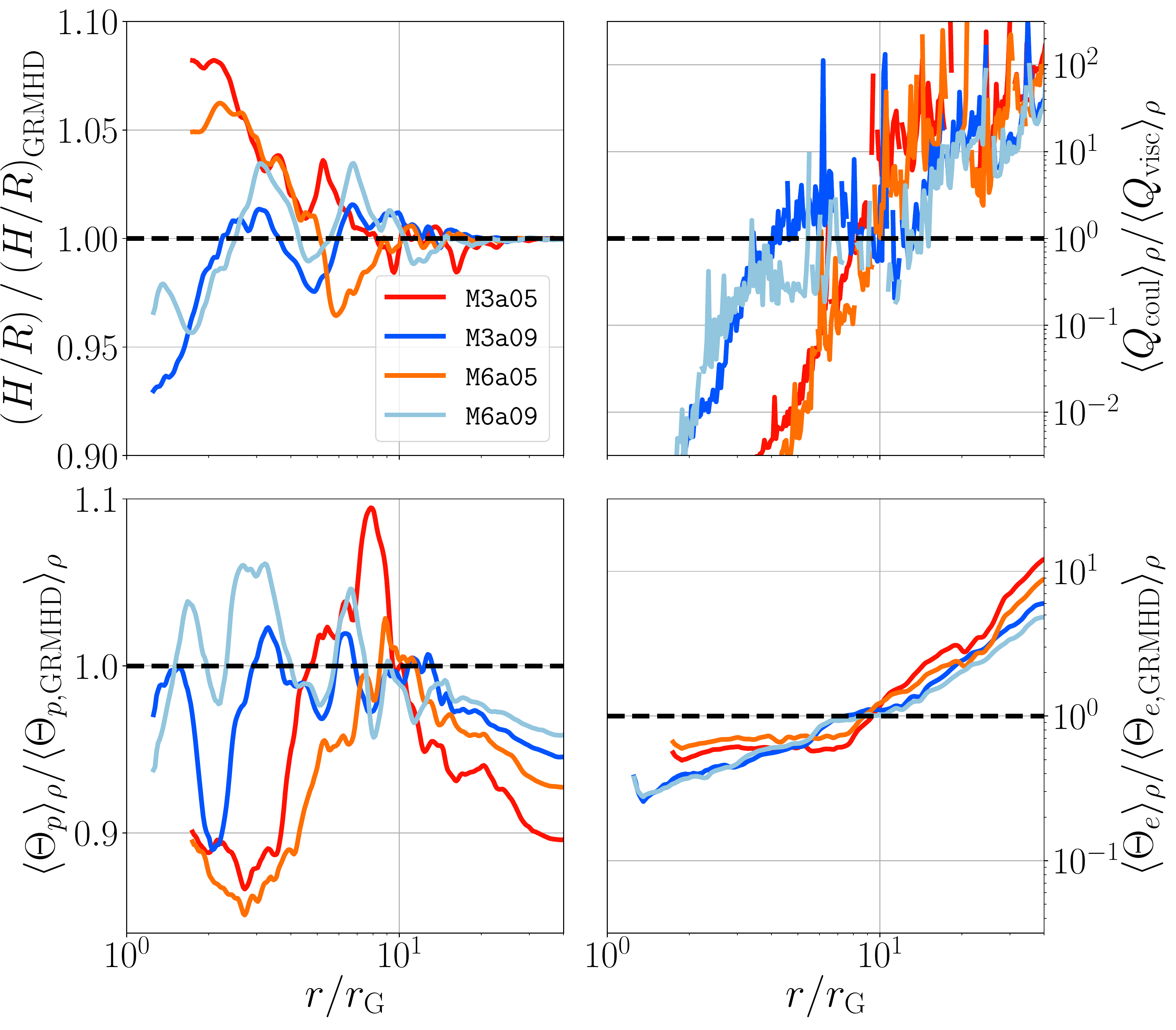}
\caption{Clockwise from top left: $H/R$, ratio of Coulomb to dissipative heating, dimensionless electron temperature $\Theta_e$, and dimensionless proton temperature $\Theta_p \equiv k_B T_p / (m_p c^2)$. $H/R$, $\Theta_e$, and $\Theta_p$ are shown relative to the same quantities from equivalent nonradiative GRMHD simulations. Dashed lines correspond to no change between the radiative and nonradiative models. $Q_{\mathrm{coul}}/Q_{\mathrm{visc}}$ is boxcar averaged for clarity.}
\label{fig:sadw_4}
\end{figure}

Figure \ref{fig:sadw_4} shows time-averaged radial profiles from all simulations. To interpret the role of radiation physics, we have performed {\it nonradiative} (hereafter `GRMHD') simulations at the two spin values from our axisymmetrized 3D data. Notice that black hole mass and accretion rate have no meaning in scale-free GRMHD simulations. Except for disabling radiation and Coulomb physics, all simulation properties are identical for GRMHD and GRRMHD simulations at each spin. 

The top left panel of Figure \ref{fig:sadw_4} shows the scaleheight $H/R$ for each simulation relative to the equivalent GRMHD simulation. Evidently $H/R$ changes by $\lesssim 5\%$, and only near the black hole. Radiative losses do not change the accretion flow geometry. In particular, there is no development of a thin, radiatively efficient disk anywhere in our simulations.

The top right panel of Figure \ref{fig:sadw_4} shows the ratio of Coulomb and turbulent heating for electrons. For $r \gtrsim 10 r_{\rm G}$, Coulomb interactions dominate the electron heating. Note that for simulations with $a_{\star} = 0.9375$, Coulomb heating over (plasma $\beta$-dependent) dissipative heating is an order of magnitude higher than in simulations with $a_{\star} = 0.5$ for $r \lesssim 10 r_{\rm G}$. This is likely a consequence of the change in radius of the innermost stable circular orbit with spin, which in turn influences the accretion disk magnetization out to some radius.

Figure \ref{fig:sadw_4} also shows proton and electron temperatures relative to their respective GRMHD simulations. Proton temperatures decrease relative to GRMHD values by $\lesssim 10\%$ very close to the black hole, and $\lesssim 10\%$ at larger radii. For $r \gtrsim 10 r_{\rm G}$, the small change in $H/R$ combined with the $\sim 5-10 \%$ drop in $\Theta_p$ implies the disk is receiving more electron pressure support at these radii in GRMHD models. 

Electron temperatures vary significantly between radiative and GRMHD models. Near the black hole, mean electron temperatures are a factor $\sim 2-3$ lower than in similar nonradiative models. For $r \gtrsim 10$, however, electrons are a factor $\sim 5-10$ hotter than in similar models that neglect Coulomb coupling. Note, however, that in the absence of fully developed turbulence far from the black hole in these simulations, dissipative heating is suppressed in this region. Therefore, despite the greatly enhanced electron heating for $r \gtrsim 10$, $\langle T_p/T_e \rangle$ is always $\gtrsim 20$ in this region.

When recording superphotons at the outer radiation boundary, we record whether they have undergone scattering events. Assuming large Compton amplification factors, i.e.\ $\Theta_e \gtrsim 1$, we can calculate the Compton $y$ parameter, the relative importance of Compton scattering to photon emission. Total luminosity is related to luminosity from emission by $L \sim L_{\mathrm{em}}(1 + y)$. Table \ref{tab:models} reports ${L_{\mathrm{em}}}/{L_{\mathrm{sc}}}$, roughly $1/y$, for each model. Compton $y$ ranges from $\sim 0.6$ to $3$; Compton scattering is an important contribution to the total luminosity, and to radiative cooling.

The jet power $P_{\mathrm{J}}$ is given in Table \ref{tab:models}. All our simulations yield approximately $P_{\mathrm{J}} \sim 10^{40}-10^{41}\mathrm{~erg~s^{-1}}$. These jet powers correspond to jet efficiencies $\sim 0.02\% - 2\%$, far below the $\sim 100\%$ efficiencies seen in rapidly spinning, strongly magnetized GRMHD simulations (\citealt{Tchekhovskoy+2011}).

\subsection{Spectra}

We now return to comparing simulated and observed quantities for M87. We first consider flux across the observed electromagnetic spectrum. The millimeter flux is fixed by EHT measurements, which resolve the source to within the computational volume we consider here (\citealt{Doeleman+2012}). Lower frequency measurements are from progressively larger structures due to increasing optical depth to synchrotron self-absorption.

The next lowest-frequency data comes from IR and optical observations (\citealt{Prieto+2016}). These measurements have an angular resolution $\sim 0.15''$; for a black hole mass $M=(3.3\times10^9,6.2\times10^9)$; this corresponds to a radius $r = (7.7\times10^4,4.1\times10^4)r_{\rm G}$, outside our simulation volume. X-ray data has a slightly lower resolution, $\sim 0.4''$. With these angular resolution, flux from the brightest M87 jet knot, HST-1, is excluded.

We cannot guarantee that we are capturing the emission region for frequencies outside of $\sim 230 $ GHz. Spectra from our \ebhlight{} models are thus best interpreted as lower limits on the emission; while larger radii may contribute to the luminosity, optical depth is low past the synchrotron peak. $R_{95}$, the radius inside of which $95\%$ of the luminosity is generated, for our models is always contained by the radiative region of each simulation. However, our models may not be in equilibrium at large radius, and do not include bremsstrahlung emission, which may contribute, especially in the X-ray, far from the black hole. 

\begin{figure}
\centering
\includegraphics[width=\figwidth]{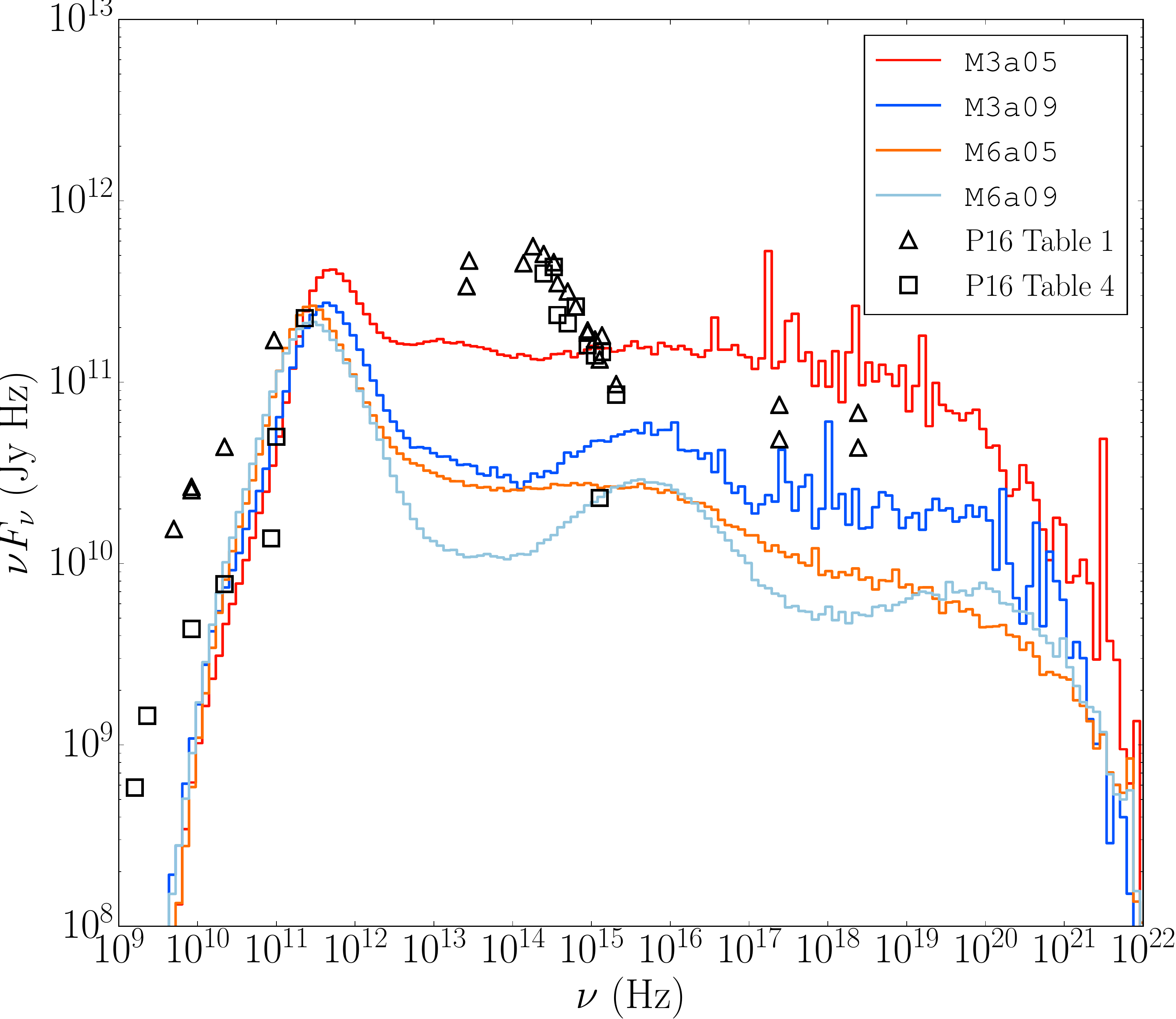}
\caption{Time-averaged face-on spectra for all models. Data points taken from quiescent period measurements in \cite{Prieto+2016} (P16). Triangles show observations for angular resolutions $\leq 0.4''$, while squares show observations for angular resolutions $\leq 0.15''$.}
\label{fig:nuFnu}
\end{figure}

Spectra are taken directly from \ebhlight{} simulations by recording superphotons crossing the outer radial radiation boundary, binned in elevation $\theta$. For M87, we consider the bin closest to the polar axis, corresponding to angles $\lesssim 35 \degree$ from the polar axes, averaged about the midplane. The time-averaged result for all models is shown in Figure \ref{fig:nuFnu}. Also shown are quiescent state observations from \cite{Prieto+2016}, given for $0.15''$ and $0.4''$ maximum angular resolutions. 

While all models recover similar millimeter slopes broadly consistent with high angular resolution measurements, no model reproduces both the optical/IR and X-ray data simultaneously. In this regard M3a09 is the most successful, producing the most flux in both bands without excluding itself. However, it still underproduces the IR by over an order of magnitude. We find agreement between simulation and high-resolution observations down to $\sim 43~{\rm GHz}$, but at lower frequencies our models underpredict the observed flux.
%While all models recover similar mm slopes broadly consistent with high angular resolution measurements, no model reproduces both the optical/IR and X-ray data simultaneously. In this regard the best performer is {\tt M3a09}, which only slightly underproduces both IR and X-ray data (thereby not excluding itself). We find agreement between simulation and high-resolution observations down to $\sim$ 43 GHz, but at lower frequencies our models underpredict the flux.

Our models do not produce much $\gamma$-ray flux, which may in any case be dominated by emission from HST-1. In addition, we do not include nonthermal electrons; a power-law tail extending to high electron energies may be responsible for extreme Compton scattering events and higher energy synchrotron emission.

\subsection{Imaging}
\label{sec:imaging}

We used escaping superphotons binned in angle to evaluate $F_{\nu,230~\mathrm{GHz}}$ for the purpose of choosing $\dot{m}$ to recover the EHT flux (\citealt{Doeleman+2012}). Images, however, are created with post-processed ray tracing along particular lines of sight (\citealt{Noble+2007}). Throughout, for imaging we set $\theta$ to either $20 \degree$ or $160 \degree$. Images are calculated with $1024\times1024$ pixels and a $70~r_{\rm G}$ field of view. We adopt a position angle, measured counterclockwise from the vertical direction in millimeter images, of 288$\degree$ \citep{Reid+1982}. For generating images, we arbitrarily choose a timeslice and $\theta = (20\degree,160\degree)$ for each simulation at which imaging-derived flux agrees with the EHT flux to within a few percent. These times and $\theta$ are given in Table \ref{tab:ipoledata}.

\begin{figure}
\centering
\includegraphics[width=\figwidth]{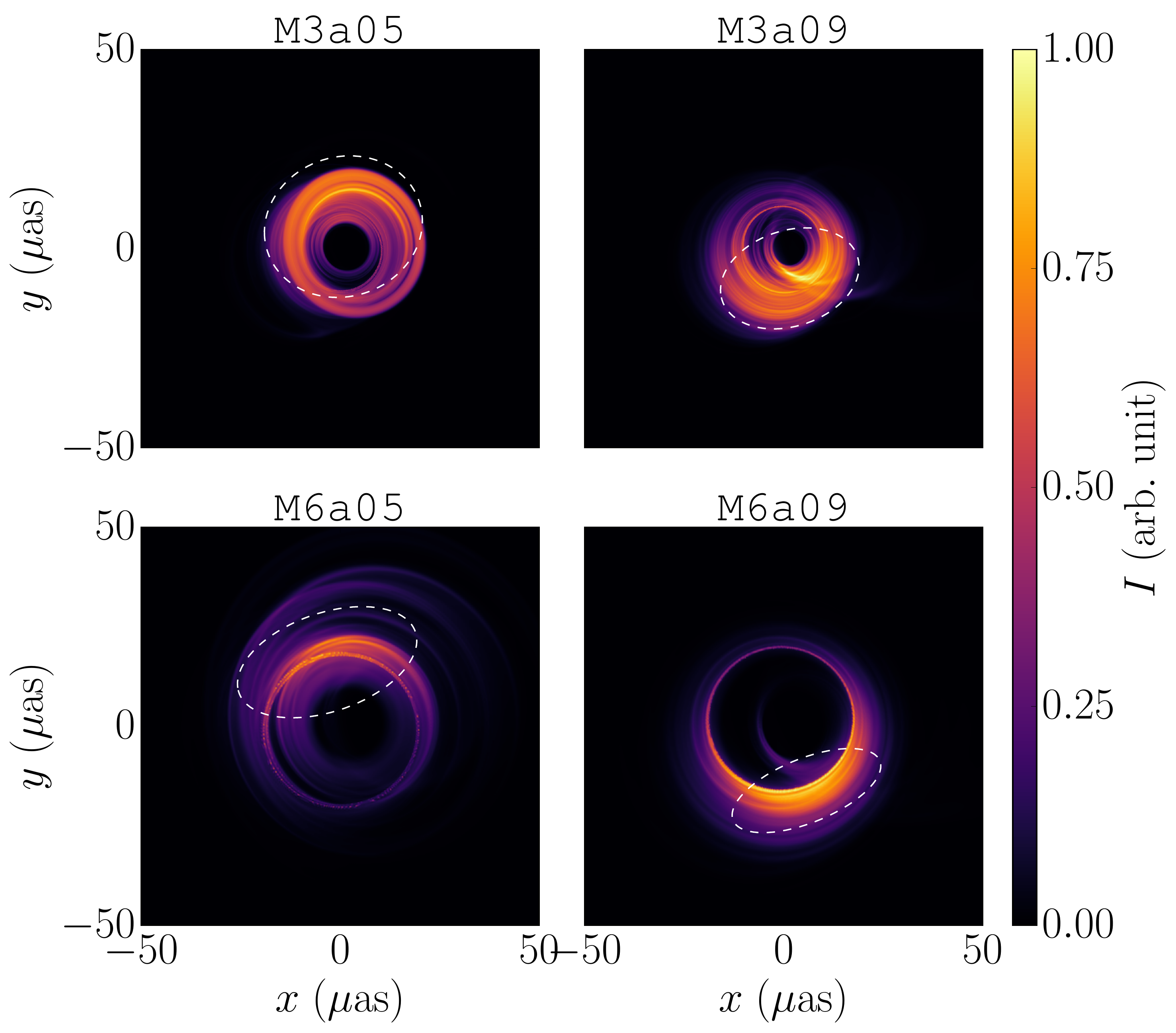}
\caption{230 GHz images from each model, evaluated at times for which the image-derived flux is nearly the value measured by the EHT (0.92 Jy; \citealt{Doeleman+2012}). Color scheme is common to all panels. Also shown are $1/e$ contours of the best-fit 2D Gaussians.}
\label{fig:mm_images}
\end{figure}

230 GHz images are shown in Figure \ref{fig:mm_images}. The size of the event horizon on the sky (the black hole shadow) is proportional to the black hole mass. Note that the relative brightness of the upper or lower half-plane (set by our choice of $\theta$) is not a prediction of our model; the orbital angular velocity of the accretion disk may be pointed either toward or away from the observer.

\begin{figure}
\centering
\includegraphics[width=\figwidth]{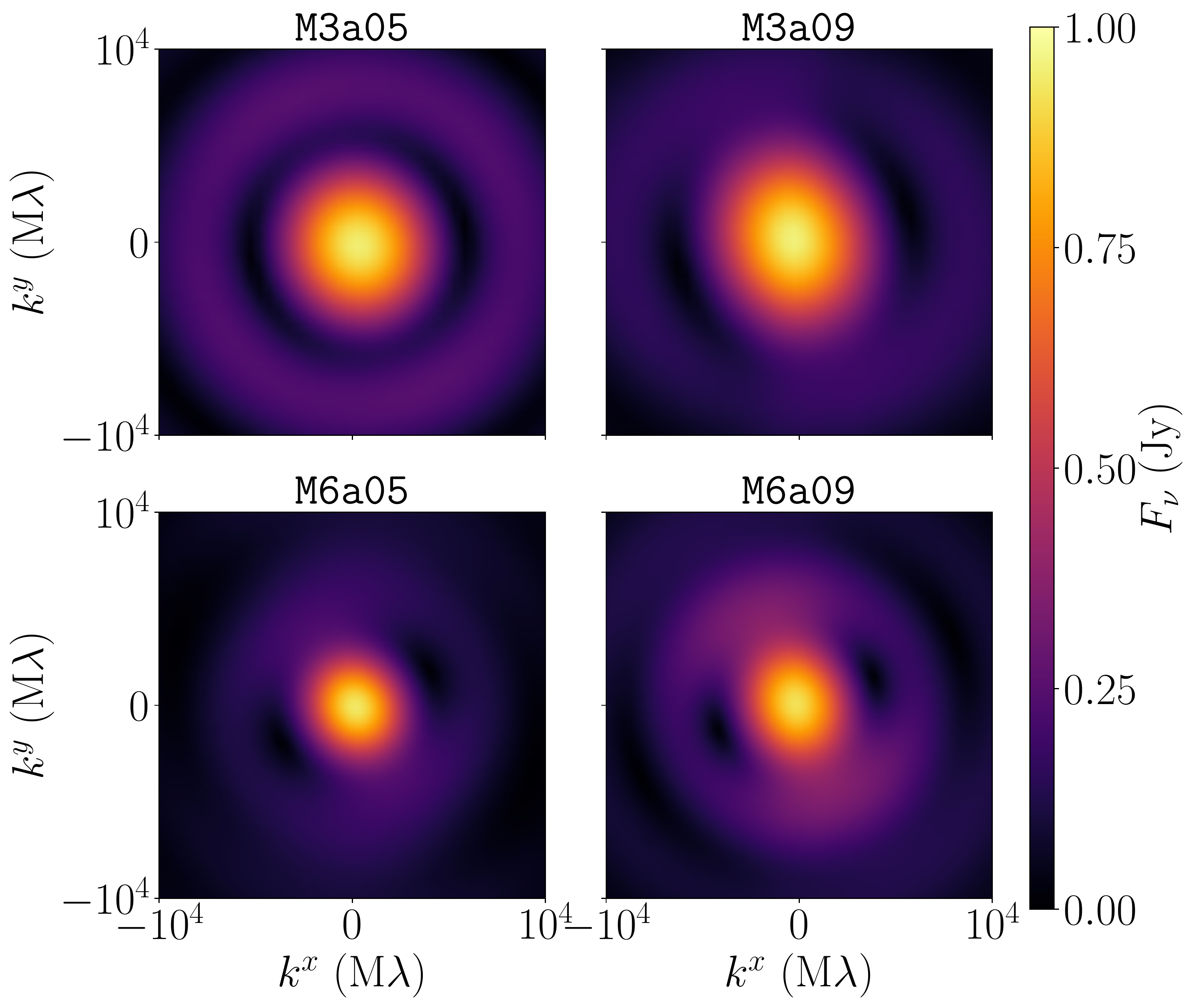}
\caption{Visibilities calculated from 230 GHz images. Images from high-mass black holes are somewhat larger on the sky, while those form low-mass black holes are more rotationally symmetric.}
\label{fig:mm_visibilities}
\end{figure}

\begin{table*}[ht]
\centering
\begin{tabular}{ccccccc}%{p{0.25\linewidth}p{0.25\linewidth}p{0.25\linewidth}}
\hline
Label & $t/M$ & $\theta$ & $f_{\mathrm{CJ}}$ & $\sigma_{\mathrm{G,maj}}$ ($\mu$as) & $\sigma_{\mathrm{G,min}}$ ($\mu$as) & $\dot{m}^{1/2} m^{-1/2} \propto \nu_s$\\
\hline
{\tt M3a05} & $850$ & $20 \degree$ & 25\% & 14.2 & 12.4 & $8.2\times10^{-8}$ \\
{\tt M3a09} & $700$ & $160\degree$ & 70\% & 12.6 & 8.5  & $5.0\times10^{-8}$ \\
{\tt M6a05} & $705$ & $20\degree$  & 34\% & 16.6 & 8.7  & $3.9\times10^{-8}$ \\
{\tt M6a09} & $745$ & $160\degree$ & 81\% & 14.1 & 5.7  & $2.9\times10^{-8}$ \\
\hline
\end{tabular}
\caption{Chosen times, inclination angles, counterjet fractions, standard deviations along major and minor axes of 2D Gaussian fits to millimeter images, and approximate relative synchrotron frequencies (see Section \ref{sec:imaging}).}
\label{tab:ipoledata}
\end{table*}

We calculate the contribution from the counterjet by setting emissivities (but not absorptivities and rotativities) to zero above and below the disk midplane (relative to the observer) to capture counter and forward jet contributions, respectively. The ratio of intensity produced below the midplane to total intensity gives the counterjet fraction $f_{\mathrm{CJ}}$, which is also given in Table \ref{tab:ipoledata}. Counterjet fraction increases with both mass and spin. 

Increased black hole spin increases the counterjet fraction in our models by a factor $\sim 3$. The sense of this effect is expected; higher spin means more emission at smaller $r$, where gravitational lensing is stronger. Our $a_{\star} = 0.5$ models show only a $\sim 30 \%$ counterjet contribution to millimeter flux. In agreement with \cite{Dexter+2012}, who study an $a_{\star} = 0.92$ GRMHD model of M87, our high-spin models are counterjet dominated. Increased black hole mass (equivalently, decreased accretion rate) increases the counterjet fraction by a factor $\sim 10-20\%$.

Figure \ref{fig:mm_images} also shows contours of least squares 2D Gaussian fits to the millimeter images. The major and minor axes, $\sigma_{\mathrm{G,maj}}$ and $\sigma_{\mathrm{G,min}}$, are given in Table \ref{tab:ipoledata}. These Gaussians vary from nearly circularly symmetric to heavily skewed; the eccentricity $e \equiv 1-\sigma_{\mathrm{G,min}}/\sigma_{\mathrm{G,maj}}$ varies from 0.13 for {\tt M3a05} to 0.60 for {\tt M6a09}. $e$ increases with both mass and spin, and is probably at least partially associated with enhanced counterjet fraction; offset circular bands of emission will be partially clipped by the black hole shadow when produced by the counterjet.

Images derived from {\tt M3a05} and {\tt M3a09} simulations in Figure \ref{fig:mm_images} are more rotationally symmetric than those from the higher black hole mass {\tt M6a05} and {\tt M6a09} simulations. While we cannot strongly constrain image variability in this work (we present only one image per simulation), at fixed 230 GHz flux we expect rounder images at lower black hole mass. Our synchrotron emissivity (\citealt{Leung+2011}) contains a factor $\exp (-(\nu /\nu_s)^{1/3})$, where the synchrotron frequency is $\nu_s \sim \dot{m}^{1/2} m^{-1/2}$ for both the ADAF (\citealt{NarayanYi1994}) and CDAF (\citealt{QuataertGruzinov2000}) models with constant plasma $\beta$. For our simulations (with fixed 230 GHz flux), we have, roughly, $\dot{m} \sim m^{-1}$ (see Table \ref{tab:models}); $\nu_s$ is larger in {\tt M3a05} and {\tt M3a09} than in {\tt M6a05} and {\tt M6a09} (see Table \ref{tab:ipoledata}). For larger $\nu_s$, the $\tau=1$ surface moves further out in radius, relativistic effects will be less apparent in images, and images will therefore more closely reflect the symmetries of the accretion flow.

\begin{figure}
\centering
\includegraphics[width=\figwidth]{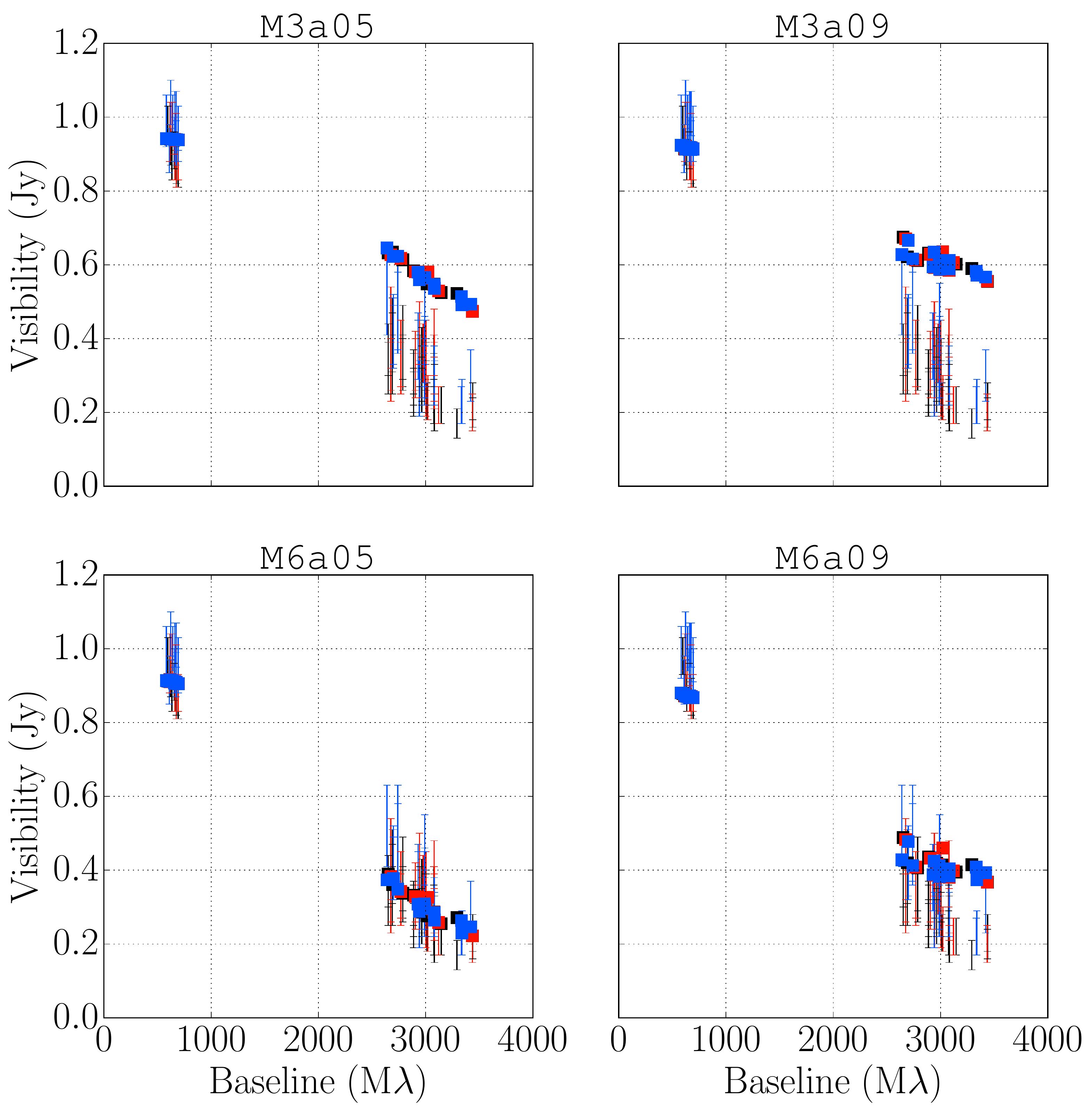}
\caption{Visibilities from EHT observations (errorbars; \citealt{Doeleman+2012}) and \ebhlight{} models (squares) for the same baselines. EHT measurements were taken over the course of three days, designated as black, red, and blue, respectively. {\tt M6a05} shows good agreement with the available data.}
\label{fig:EHT}
\end{figure}

Figure \ref{fig:mm_visibilities} shows visibility maps corresponding to the images in Figure \ref{fig:mm_images}. The black hole shadow (smaller in the low-mass models) is clearly visible in all cases as zeroes in the visibilities. However, the low-mass models are nearly rotationally symmetric, whereas the high-mass models show a strong asymmetry.

%We compare to the measured EHT visibilities in \cite{Doeleman+2012} in Figure \ref{fig:EHT} by extracting fluxes from our measured visibilities at the same baselines. These observations were taken over the course of three days (about 16 and 8.5 $t_{\rm G}$ for black hole masses $M/M_{\odot} = 3.3\times10^9$ and $6.2\times10^9$, respectively; note, however that our simulated visibilities are each calculated at a single simulation timeslice). All simulations do a reasonable job of reproducing the short baseline fluxes. The low-mass simulations, however, overproduce flux at longer baselines; these images are too small, at least for our electron physics model. {\tt M6a09} overproduces flux at the longest baselines and marginally underproduces at short baselines, while {\tt M6a05} agrees well.
We compare to the measured EHT visibilities in \cite{Doeleman+2012} in Figure \ref{fig:EHT} by extracting fluxes from our measured visibilities at the same baselines. These observations were taken over the course of three days (about 16 and 8.5 $t_{\rm G}$ for black hole masses $M/M_{\odot} = 3.3\times10^9$ and $6.2\times10^9$, respectively; note, however that our simulated visibilities are each calculated at a single simulation timeslice). These EHT baselines are clustered in short and long baseline groups; for black hole masses $M/M_{\odot} = (3.3\times10^9$, $6.2\times10^9$), the short baselines ($\sim 600 {\rm M}\lambda$) probe structures with radii $\sim( 90 r_{\rm G}, 47 r_{\rm G}$) while the long baselines ($\sim 3000 {\rm M}\lambda$) probe structures with radii $\sim( 18 r_{\rm G}, 9 r_{\rm G}$). All simulations do a reasonable job of reproducing the short-baseline fluxes. The low-mass simulations, however, overproduce flux at longer baselines; these images are too small, at least for our electron physics model. {\tt M6a09} overproduces flux at the longest baselines and marginally underproduces at short baselines, while {\tt M6a05} agrees well.

\section{Discussion}
\label{sec:discussion}

We have used \ebhlight{} simulations to study the inner region of M87's accretion flow. We have chosen mass accretion rates to recover the 230 GHz flux of \cite{Doeleman+2012}, and then tested synthetic observations from these simulations against fluxes at other wavelengths and interferometric imaging, as well as comparing measured and inferred jet powers. The low-mass, low-spin {\tt M3a05} simulation significantly overproduced X-ray emission. IR/optical emission is uniformly underproduced in our models. A strong constraint comes from the resolved millimeter visibilities of \cite{Doeleman+2012}, which are relatively insensitive to both the observational confusion and systematic uncertainties in our models. These observations are inconsistent with the compact emission from both low-mass models, {\tt M3a05} and {\tt M3a09}. The {\tt M6a09} millimeter image was slightly too small, although with only marginal significance. {\tt M6a05} generally agreed well with spectral and imaging constraints. All our models significantly underproduce the jet power inferred from observations. Despite uncertainties, multiple techniques for inferring the jet power, such as VLBI observations of the radio core (e.g.\ \citealt{de Gasperin+2012} and internal pressure in a reconfinement shock (e.g.\ \citealt{Stawarz+2006}), recover a similar jet power $P_{\mathrm{J}} \sim 10^{43} - 10^{44} \mathrm{~erg~s^{-1}}$.

Our study prefers the high-mass, low-spin model {\tt M6a05}. While one or both of {\tt M3a05} and {\tt M3a09} disagree with one or more observational constraints, {\tt M6a05} is only marginally preferable to {\tt M6a09} and our study does not provide much ability to constrain black hole spin. We now turn to future directions, particular in the context of uncertainties in this work. 

\subsection{Multifrequency Observations}

We find that high spin leads to more distinct Compton bumps, whereas low-spin models are nearly power laws between synchrotron emission and the high-frequency cutoff. This conclusion is probably sensitive to our assumption of a thermal electron distribution function. Although resolving the inner $\sim 100 r_{\rm G}$ of M87 is challenging at these wavelengths, additional frequency measurements filling out the spectrum could help constrain the spin of M87 by identifying or ruling out Compton bumps.

We now focus on the low-frequency radio slope, for which we prefer the high angular resolution values in \cite{Prieto+2016}. Below $\nu \sim 10^{10} {~\rm Hz}$, our models underpredict the observed fluxes. This may be a consequence of our limited domain size, lack of nonthermal particles, and/or issues with anomalous numerical cooling of electrons near the funnel wall. See \cite{Ressler+2017} for a discussion of the radio slope in models with self-consistent electron thermodynamics.

Our models underpredict the NIR/optical flux (apart from {\tt M3a05}, which has an inconsistent spectral shape at these frequencies). \cite{Prieto+2016} used an extrapolation method to remove background starlight. However, the observed fluxes at these frequencies could still be dominated by a radiation mechanism other than Compton upscattering in the inner $\sim 100 r_{\rm G}$. Dust emission, a stellar population, and nonthermal synchrotron could all play some role, although we will not speculate further in this work.

Our models do not produce much radiation beyond $\sim 10^{22}$ Hz. This may be a consequence of our assumption of thermal electron distributions everywhere (in contrast with ion distributions in kinetic shearing box simulations, e.g.\ \citealt{Kunz+2016}, and electron distributions in high-magnetization reconnection simulations, e.g.\ \citealt{SironiSpitkovsky2014}). M87 is a powerful emitter of TeV photons (e.g.\ \citealt{Aharonian+2006}). The high-energy emission that we observe, however, may need to originate in highly relativistic outflows (i.e.\ in the jet) to circumvent the opacity to pair production \citep{Begelman+2008}. TeV observations cannot separate the inner region of the accretion flow with jet knots, most notably HST-1. Identifying the origin of TeV photons in M87 is an important future direction.

Bremsstrahlung emission is not considered in our model; it is subdominant to synchrotron emission near the black hole. At larger radii, however, it may be a significant source of X-ray emission. Exploring a very large dynamic range in radius (and therefore time) is challenging for simulations; modeling emission from the entire region subtended by the resolution of X-ray observations is consequently difficult.

\subsubsection{Equilibration}

Axisymmetry limits the duration of our simulations. We are therefore not able to achieve inflow equilibrium at all radii for which radiation is (thermo)dynamically important; the disk structure may tend toward smaller scale heights and ion and electron temperatures at these radii. While such a change would leave 230 GHz emission and images largely unchanged, it could sap energy from the spectrum throughout the region dominated by Compton upscattering.

\subsubsection{Electron heating}

The electron temperature is simultaneously perhaps the most important and most uncertain component of our simulations. While \cite{Ressler+2015} represents a significant advance from ad hoc models, it is vulnerable to numerical challenges in accurately capturing grid-scale dissipation, and uncertainties in the underlying kinetic physics itself.

\cite{Ressler+2017} provide a discussion of difficulties in measuring dissipation due to truncation error. For one, dissipation can be either positive or negative (total energy-conserving schemes locally obey the second law of thermodynamics only to truncation error). At least for uniform low Mach number turbulence, such as generally obtained near the midplane in MRI-driven RIAF simulations, dissipation acts as heating on average. In the presence of large entropy gradients, however, dissipation can have a net cooling effect. The funnel wall is such a configuration, and our funnel wall electrons may be artificially cool. This has potential consequences for the low-frequency radio slope; artificial electron cooling will suppress low-frequency emission.

We employ the state-of-the-art electron heating fraction model of \cite{Howes2010}. For this model, \cite{Howes2011} found agreement within experimental uncertainty with the electron-ion temperature ratio in the solar wind, probably the best accessible analog of RIAFs. However, uncertainties are non-negligible, and the data do not cover the entire range of plasma $\beta$ present in our simulations. Our understanding of microscale electron heating is incomplete (e.g.\ \citealt{Rowan+2017}), and new results (as well as nonthermal electron distributions) may substantially change the results of global simulations (e.g.\ \citealt{Chael+2018}).

\subsection{Net Magnetic Flux}
\label{sec:netflux}

M87 sources a powerful relativistic jet, and such jets may be associated with black holes accreting at the MAD limit (\citealt{Narayan+2003}, \citealt{Tchekhovskoy+2011}, \citealt{McKinney+2012}). In MADs, strong vertical magnetic fields qualitatively change the accretion flow. The interchange instabilities, which govern angular momentum transport in this case, are probably inaccessible to axisymmetric fluid models, such as we study here, and hence we avoid consideration of MADs. Nonetheless, this is a viable model for the M87 accretion flow. The $\sim 100\%$ efficiencies in the electromagnetic jet luminosity would, all else being equal, bring our measured $P_{\mathrm{J}}$ in line with inferred values, as our current jet efficiencies are $\sim 0.02 \%$ and $\sim 2\%$ for the low- and high-spin models, respectively.

In our simulations the dimensionless net magnetic flux through the black hole $\phi \sim 4-7$, whereas for MADs $\phi \sim 50$ (\citealt{Tchekhovskoy+2012}). The funnels in our models have relatively narrow opening angles compared to MAD simulations (e.g.\ \citealt{Tchekhovskoy+2011}). The consequences of a wider jet in our model, especially for coronal electron temperatures, is uncertain (although see \citealt{Ressler+2017} for a semi-MAD calculation with electron heating). 

Whether MAD models with self-consistent radiative cooling are a viable alternative for M87 is an interesting question we plan to explore in future work. Along with spectra, MADs may exhibit quite different variability and polarization (e.g.\ \citealt{Gold+2017}). We caution, however, that the greater magnetization in MADs presents steeper numerical challenges to conservative GRMHD schemes, particularly when evaluating fluid temperatures. This problem is compounded when using the \cite{Ressler+2015} method for electron heating. Improvements to existing numerical GRMHD methods may be required to pursue this question with satisfactory accuracy. 

\subsection{Variability}

Our axisymmetric model has limited duration and probably overestimates variability. Additionally, nonaxisymmetric fluctuations may imprint characteristic frequencies onto light curves, e.g.\ \cite{Dolence+2012} and \cite{ShcherbakovMcKinney2013}. Hence, we leave the study of variability to future work. Studies of variability in 3D GRRMHD simulations of M87, and RIAFs subject to cooling more generally, are a promising future direction for constraining accretion flows.

After compiling separate observed spectra for M87 in quiescent and active states, \cite{Prieto+2016} argue that the spectral shape seems independent of state; the entire spectrum simply shifts up or down. Given that the accretion rates we study already show the effects of radiative cooling, increasing $\dot{m}$ to match the active state of that in \cite{Prieto+2016}, assuming an increase in accretion rate in the source is responsible for the outburst, would presumably serve to increase radiative cooling. Cooling tends to alter spectral shape; for example, when distinct Compton bumps are present, their separation is $\sim A$, the amplification factor. Requiring that a single model recovers both quiescent and active spectra could act as a powerful discriminant in the future.

\section{Conclusion}
\label{sec:conclusion}

We have presented two-temperature GRRMHD models of the inner accretion flow of M87. Along the way, we considered the interplay of dissipative heating, Coulomb coupling, and radiative cooling in RIAFs at $\dot{M}/\dot{M}_{\rm Edd} \sim 10^{-5}$. We found that Compton $y$ parameters $\sim 1$ for these models, consistent with previous estimates. We find that Coulomb heating dominates dissipative heating for electrons for $r \gtrsim 10 r_{\rm G}$. We have demonstrated that radiative cooling is important for the inner region of the M87 accretion flow in our model.

For black hole masses bracketing the observationally preferred values and high and low black hole spins, we have derived synthetic observations of spectra and 230 GHz images. Acknowledging uncertainties in our chosen net magnetic field and electron heating model, we exclude a low black hole mass, $M/M_{\odot} = 3.3\times10^9$, through radio image sizes, and the low-mass, low-spin model through overproduction of X-rays. $M/M_{\odot} = 6.2\times10^9$ simulations satisfy radio/IR/X-ray emission and image size. However, jet power is always a factor $10^2-10^3$ lower than previously inferred values. This is probably a consequence of the absence of a strong large-scale poloidal field in our initial conditions.

Our model is axisymmetric, which not only limits our time integration window but also renders variability information unreliable. Similar modeling in three spatial dimensions is a critical future direction, albeit much more expensive, especially given our procedure for determining the optimal accretion rate through a series of simulations.

\acknowledgments
It is a pleasure to thank J. Dexter, M. Moscibrodzka, A. Tchekhovskoy, and Fu-Guo Xie for useful discussions. Work at Los Alamos National Laboratory was done under the auspices of the National Nuclear Security Administration of the U.S.\ Department of Energy. S.M.R.\ is supported in part by the NASA Earth and Space Science Fellowship. J.D.\ acknowledges support from the Laboratory Directed Research and Development program at Los Alamos National Laboratory. C.F.G.\'s work was also supported in part by a Romano Professorial Scholar appointment. This work was supported in part by NSF grants AST 13-33612, AST 1715054, {\it Chandra} theory grant TM7-18006X from the Smithsonian Institution, and a Simons Investigator award from the Simons Foundation. This work was made possible by computing time granted by UCB on the Savio cluster. This work benefited from the Extreme Science and Engineering Discovery Environment (allocation TG-AST170024), which is supported by National Science Foundation grant number ACI-1053575. This research used resources provided by the Los Alamos National Laboratory Institutional Computing Program, which is supported by the U.S. Department of Energy National Nuclear Security Administration under Contract No.\ DE-AC52-06NA25396. This article has been assigned a LANL document release number LA-UR-18-23675.

\software{{\tt ebhlight} \cite{Ryan+2017}, {\tt bhlight} \cite{Ryan+2015}}

\newpage
%\bibliographystyle{apj}
%\bibliography{local}

\end{document}